\documentclass[twocolumn,preprintnumbers,amsmath,amssymb,floatfix,showpacs,apl]{revtex4}

\usepackage{graphicx}
\usepackage{dcolumn}
\usepackage{bm}

\begin{document}

\title{A spintronic source of circularly polarized single photons}

\def\APH{Institut f{\"u}r Angewandte Physik and DFG Center for
Functional Nanostructures (CFN), Karlsruhe Institute of Technology
(KIT), 76131 Karlsruhe, Germany}

\author{Pablo Asshoff}
\email[]{pablo.asshoff@kit.edu}
\affiliation{\APH}
\author{Andreas Merz}
\affiliation{\APH}
\author{Heinz Kalt}
\affiliation{\APH}
\author{Michael Hetterich}
\affiliation{\APH}

\begin{abstract}

We present a spintronic single photon source which emits circularly
polarized light, where the helicity is determined by an applied
magnetic field. Photons are emitted from an InGaAs quantum dot
inside an electrically operated spin light-emitting diode, which
comprises the diluted magnetic semiconductor ZnMnSe. The circular
polarization degree of the emitted light is high, reaching 83\,\% at
an applied magnetic field of~2\,T and~96\,\% at~6\,T.
Autocorrelation traces recorded in pulsed operation mode prove the
emitted light to be antibunched. The two circular polarization
states could be used for representing quantum states $\mid0>$ and
$\mid1>$ in quantum cryptography implementations.

\end{abstract}



\maketitle


The most established materials for enabling spin-polarization of
electronic carriers are ferromagnets, half-metals and magnetic
semiconductors~\cite{TrinitySemiconductorsHalfmetalsFerromagnets}.
Excellent spin-injection efficiencies into adjacent semiconductor
structures have been achieved with magnetic
semiconductors~\cite{EuOSchlom, MacDonaldDMS}, resulting from their
similar physical properties and the high interface quality. Due to
this outstanding performance, it makes magnetic semiconductors a
material of choice for designing an electrically operated light
source for photons with defined helicity. A spin-polarized
electrical current created by the magnetic semiconductor and
injected into a quantum dot (QD) will result in emission of
circularly polarized light, assuming the ideal case that the spin
state of the carriers is preserved until optical
recombination~\cite{HetterichPss}. Furthermore, in order to generate
single photons on demand, when a voltage pulse is applied to the
device, the light emitted will need to be
antibunched~\cite{ElectricalSinglePhotonSourceScience}. A classical
structure for the implementation of these design prerequisites is a
spin light-emitting diode (spin-LED), which usually serves as test
device for determining the spin-polarization degree of injected
carriers. However, as a single photon source with $\sigma^+$- or
$\sigma^-$-polarized emission, it could be employed for securely
coding and transmitting data for applications in quantum
cryptography. In this paper we will show that such a spin-LED can be
realized, which emits circularly polarized light exhibiting very
high polarization degrees, that the helicity of the light can be
controlled by the direction of an external magnetic field, and that
by applying voltage pulses antibunched light is generated.

\begin{figure}[b] \centering
\includegraphics[clip=true,keepaspectratio=true,width=0.58\linewidth]{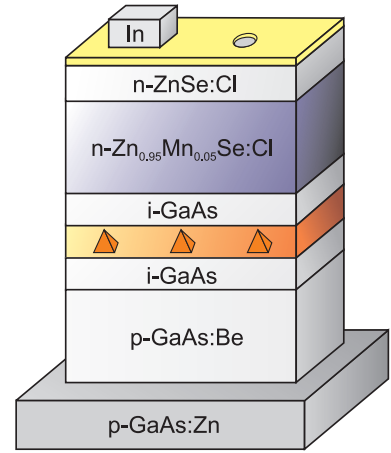}
\caption{\label{scheme}Schematic of the spintronic single photon
emitting diode. InGaAs QDs are sketched in the optically active
region and an aperture on top of the heterostructure is shown.}
\end{figure}

The basic design layout of the specific sample presented is sketched
in Fig.~\ref{scheme}, where the most important functional elements
are marked in color. The single photon source is a single QD
emitting circularly polarized photons at~$\sim1.348$~eV, which is
located in the optically active region (orange layer) of a spin-LED
with a surface area of $(400 \ \mu \mathrm{m})^2$. The diluted
magnetic semiconductor (DMS) ZnMnSe is integrated into the
heterostructure (blue layer), it renders traversing electrons
spin-polarized when a magnetic field is applied. This is due to
electrons relaxing into the energetically lower of the two
spin-split conduction bands, which are separated through a giant
Zeeman splitting~\cite{FurdynaDMS}. Thus, applying a voltage across
the spin-LED results in spin-polarized electrons reaching the QD,
where they recombine with unpolarized holes injected from the bottom
part of the spin-LED. Due to optical selection rules, these
transitions can only take place under the emission of circularly
polarized light. A submicron aperture on top of the heterostructure
(within the yellow top layer) helps to minimize spurious emission
from other nearby QDs.

Fabrication and processing of the sample were carried out as
follows. A GaAs:Zn(001) wafer ($p\sim 1\times 10^{19}$ cm$^{-3}$)
was used as substrate. On the substrate a $\sim$\,500\,nm layer of
GaAs:Be ($p\sim 1\times 10^{19}$ cm$^{-3}$) was grown using a III--V
molecular-beam epitaxy (MBE) facility, followed by 100\,nm $i$-GaAs,
the InGaAs QDs / wetting layer (WL)~(see Ref.~\cite{PassowQDgrowth}
for details) and a 25\,nm thick \mbox{$i$-GaAs} spacer. Plan-view
transmission electron microscopy revealed a QD sheet density of
$\sim5 \times 10^{10}$~cm$^{-2}$. From optical characterization we
found that the QDs have a negligible in-plane asymmetry. The
heterostructure was then transferred to a second MBE facility,
designed for the growth of II--VI materials, where 750\,nm of the
DMS Zn$_{0.95}$Mn$_{0.05}$Se:Cl ($n \sim 10^{18}$\,cm$^{-3}$)
followed by a 200\,nm layer of ZnSe:Cl ($n = 5 \times
10^{18}$\,cm$^{-3}$) were deposited. The latter layer improves the
ohmic contact to the subsequently evaporated In contact pad. Then, a
thin gold layer was thermally evaporated wherein apertures were
defined by electron beam lithography. Finally, optical lithography
was employed to obtain square-shaped spin-LEDs.


We now analyze our device optically. First, we show that the QD
emission is spectrally isolated and determine its polarization
degree for varying magnetic fields. As measure of the photon
polarization state we define the circular polarization degree
$P_C=(I_{\sigma^+}-I_{\sigma^-}) / (I_{\sigma^+}+I_{\sigma^-})$,
with $I_{\sigma^{+(-)}}$ denoting the intensity of
$\sigma^{+(-)}$-polarized light. $P_C$ has been shown to correspond
to the spin-polarization degree of the injected electrons, since
when electrons and holes recombine in the QD, only heavy hole states
contribute to optical transitions~\cite{HetterichPss}. Experiments
were carried out in a magneto-optical cryostat with the sample
temperature set to $T=5$~K. The magnetic field was applied in
Faraday geometry ($\mathbf{k} \| \mathbf{B}$) and swept from -6\,T
to +6\,T. A 60\,$\times$ microscope objective mounted inside the
cryostat allowed to guide light from one aperture of the device,
positioned by piezoelectric actuators at the focal point, to the
outside of the cryostat. To differentiate between $\sigma^+$- and
$\sigma^-$-polarized photons, a quarter-wave plate and a linear
polarizer were inserted in the beam path. This array was arranged
such that either $\sigma^+$- or $\sigma^-$-polarized photons were
remaining after having passed the filters. Light with intensity
$I_{\sigma^{+/-}}$ was then focused into a multi-mode fiber and
transmitted to a double spectrometer with two 1200~grooves/mm
gratings, where it was spectrally dispersed. A charge-coupled device
was used to detect the diffracted light. Using this setup, a
spectral resolution in the order of $20 \ \mu$eV per pixel was
achieved.

\begin{figure}[t] \centering
\includegraphics[clip=true,keepaspectratio=true,width=0.7\linewidth]{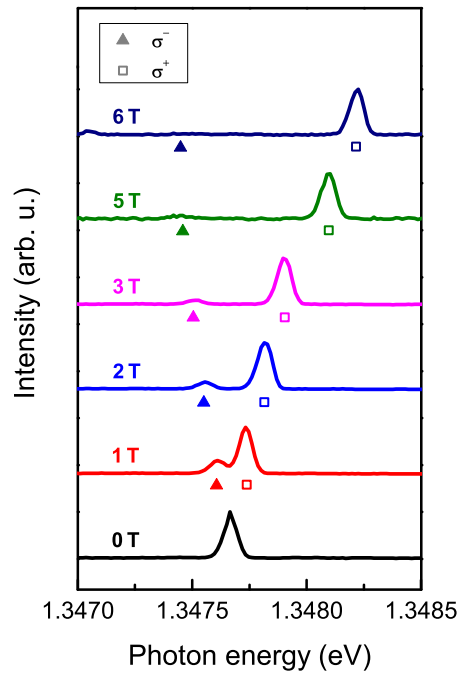}
\caption{\label{spectra}Spectrally resolved emission from the QD
single photon source at $T=5$~K and $B\geq0$~T during excitation
with a continuous current of 0.29~mA. As the magnetic field is
increased, emission splits into two components from the spin-up and
spin-down levels, which are indicated by the markers.}
\end{figure}

\begin{figure}[b] \centering
\includegraphics[clip=true,keepaspectratio=true,width=1.0\linewidth]{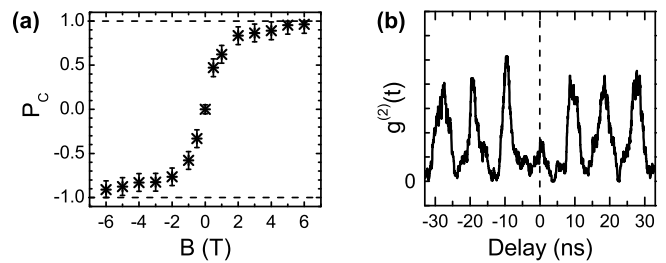}
\caption{\label{maingraph}Main characteristics of the spintronic
single photon source ($T=5$~K): (a) The average circular
polarization degree $P_C$ of the emitted photons from the QD (with a
continuous current of 0.29~mA) reaches high values. Typical error
bars obtained from repeated measurements are indicated. (b)
Autocorrelation measurement of $\sigma^+$-polarized photons emitted
from the QD at an applied field of +2~T. The sample is excited with
sub-nanosecond pulses from a fast electrical pulser. From the zero
delay peak area we obtain a value of $g^{(2)}(0)=0.38$ for the
correlation signal, giving evidence of single photon emission.}
\end{figure}

In Fig.~\ref{spectra} the obtained electroluminescence spectra of
the selected QD during continuous excitation with a current of
0.29~mA are shown. Selectively recorded $\sigma^+$- and
$\sigma^-$-polarized spectral components were integrated in this
graph. The degeneracy of the spin-up and spin-down sublevels in the
QD is lifted when a magnetic field is applied (Zeeman splitting),
shifting the $\sigma^{+}$/$\sigma^{-}$-polarized excitonic emission
from the spin-down/spin-up conduction band sublevel to higher/lower
energies. The spectral position of the emission from the spin
sublevels ($\uparrow, \downarrow$) shifts according to $E_{\uparrow,
\downarrow}= E_0 \mp \frac{1}{2} g_{exc} \mu_{B} B + \gamma B^2$
with $g_{exc}=2.22$, where the Zeeman splitting and the diamagnetic
shift are considered. The upper spin subband is filled with
spin-down electrons from the DMS, which occupy a higher energy state
in the QD~\cite{HetterichPss}. Therefore, spin relaxation within the
QD would populate the lower spin sublevel. The spectra show that the
net spin-polarization $P_C$ increases strongly (corresponding to an
increase of the normalized value of $I_{\sigma^+}-I_{\sigma^-}$)
along with magnetic field to high values. At high magnetic fields,
it can be seen that the emission is originating preferably from the
upper spin subband and the minority component vanishes almost
completely at $B=+6$~T. As the upper spin sublevel is populated, it
is clear that spin-polarization originates from the
DMS~\cite{footnote1}. Significant circular polarization degrees
$P_C$ are already reached at much lower fields. The corresponding
circular polarization degrees are shown in Fig.~\ref{maingraph}(a),
where also results for reversed magnetic fields are given. It can be
seen that without an applied magnetic field the DMS does not
spin-polarize electrons, leading to a net spin-polarization of the
electrons recombining with holes in the QD of zero. This is due to
the paramagnetic nature of the specific DMS used here, which
requires an external magnetic field to effectively polarize the
electron spins in the conduction band~\cite{FurdynaDMS}. As noticed
before, when the magnetic field is turned on, the circular
polarization degree increases rapidly. At an applied field of +2~T a
circular polarization degree of 83\,\% is observed, while at +6~T it
approaches 96\,\%, indicating emission is almost completely
$\sigma^{+}$-polarized. When the magnetic field is reversed, the
circular polarization degree changes sign, but reaches the same
magnitude. This corroborates that the high values of $P_C$ are not
caused by experimental artifacts. With reversed field, emitted
photons now are predominantly $\sigma^{-}$-polarized. The desired
circular polarization can thus be chosen by the direction of the
applied magnetic field. Furthermore, the value of $P_C$ can be
further increased by reducing the current density through the
spin-LED. For instance, with $I=0.26$~mA and $B=+6$~T we obtained
$P_C =$~98\;\%. However, along with a lower current density the
overall emission intensity is further reduced and close to the
detection limit of the spectroscopic setup. During pulsed operation
of the device, the circular polarization degree should exhibit a
time-dependency, and its time-averaged value should be higher than
in continuous operation~\cite{AsshoffSpinLEDPulsed}.

In a second experiment, used to show antibunching of the circularly
polarized photons from the QD, the previously described setup was
modified. The magnetic field inside the magneto-optical cryostat was
set to $B=+2$~T. A pulse generator excited the spin-LED with
sub-nanosecond electrical pulses. The pulses had an offset slightly
below the threshold voltage of the spin-LED, a repetition frequency
of 106.5~MHz and a nominal peak width of 0.5~ns. The signal was
transmitted to the spin-LED with a low-temperature coaxial cable. To
reduce spurious effects from impedance mismatch, parallel to the
spin-LED we mounted a 1 nF capacitor in series with a 50 $\Omega$
resistor. Emitted light was again collected by a 60\,$\times$
microscope objective and left the cryostat through an optical window
as parallel beam. In the optical detection path, the quarter-wave
plate and the linear polarizer were removed to allow more of the
signal to pass. After being transmitted to the spectrometer by a
fiber, the light left the double spectrometer at the first exit
port, thereby being diffracted by only one of the two 1200
grooves/mm gratings. The spectrometer was adjusted such that only
the more intensive $\sigma^{+}$-polarized excitonic emission line
emitted from the QD reached the exit port. Then, the light was
collimated into a 1$\times$2 multi-mode fiber optical splitter with
a nominal splitting ratio of 50/50. The splitted signals were
outcoupled from the two terminal fiber connectors and focused onto
single photon counting modules (nominal photon detection efficiency
of 33\,\% at relevant wavelength, 25 dark counts/s). This
fiber-based Hanbury-Brown Twiss setup was chosen to prevent false
coincidence signals from cross-talk effects~\cite{BeckerHicklBook}.
Antibunching was detected by measuring coincidence counts at zero
time delay between the two single photon counters. The second order
correlation function $g^{(2)}(t)$ obtained from this experiment is
shown in Fig.~\ref{maingraph}(b). From the value of the zero delay
peak area $g^{(2)}(0)=0.38$ it clearly results that the emission is
nonclassical, proving the layer sequence, compositions and
thicknesses of the design are suitable for obtaining single photons
by applying voltage pulses.

Although during pulsed excitation WL emission adds to the overall
sample emission to a larger extent than under continuous current
excitation, WL states do contribute only faintly to the spectrally
filtered signal reaching the single photon counters. This is due to
the WL emission being centered at 1.40\,eV, such that the QD
emission lines are only marginally influenced. We assume that the
value of the zero delay peak area can be significantly reduced by
varying the excitation parameters, e.g., by depleting the spin-LED
of carriers with a short negative voltage pulse immediately after
the excitation to further suppress multiphoton emission. Evidence of
this for an electrically driven single photon source with an InAs QD
was reported in Ref.~\cite{StarkShiftSinglePhotonSource}.


To conclude, we have demonstrated an electrically operated
spintronic device based on a magnetic semiconductor which emits
single photons with a selected circular polarization degree. The
measured circular polarization degrees are very high, indicating the
effectivity of our approach. The helicity of the emitted photons can
be controlled by the direction of an externally applied magnetic
field. The presented design is limited by the magnetic field and
temperature requirements. Further development of ferromagnetic
semiconductor materials may lift these requirements. However, the
material used in this study may serve as a model system
demonstrating a possible design, and suggests that magnetic
semiconductors are most promising for the realization of light
sources for single photons with defined helicity.


This work has been performed within project A2 of the DFG Research
Center for Functional Nanostructures (CFN). It has been further
funded by a grant from the Ministry of Science, Research and the
Arts of Baden-W{\"u}rttemberg (Az.: 7713.14-300). We acknowledge
financial support from the Karlsruhe School of Optics and Photonics
(A.M. and P.A.) and the Karlsruhe House of Young Scientists (P.A.).

\bibliographystyle{apsrev}

\begin{thebibliography}{11}
\expandafter\ifx\csname
natexlab\endcsname\relax\def\natexlab#1{#1}\fi
\expandafter\ifx\csname bibnamefont\endcsname\relax
  \def\bibnamefont#1{#1}\fi
\expandafter\ifx\csname bibfnamefont\endcsname\relax
  \def\bibfnamefont#1{#1}\fi
\expandafter\ifx\csname citenamefont\endcsname\relax
  \def\citenamefont#1{#1}\fi
\expandafter\ifx\csname url\endcsname\relax
  \def\url#1{\texttt{#1}}\fi
\expandafter\ifx\csname urlprefix\endcsname\relax\def\urlprefix{URL
}\fi \providecommand{\bibinfo}[2]{#2}
\providecommand{\eprint}[2][]{\url{#2}}

\bibitem[{\citenamefont{Coey and
  Sanvito}(2004)}]{TrinitySemiconductorsHalfmetalsFerromagnets}
\bibinfo{author}{\bibfnamefont{J.~M.~D.} \bibnamefont{Coey}} \bibnamefont{and}
  \bibinfo{author}{\bibfnamefont{S.}~\bibnamefont{Sanvito}},
  \bibinfo{journal}{J. Phys. D: Appl. Phys.} \textbf{\bibinfo{volume}{37}},
  \bibinfo{pages}{988} (\bibinfo{year}{2004}).

\bibitem[{\citenamefont{Schmehl et~al.}(2007)\citenamefont{Schmehl,
  Vaithyanathan, Herrnberger, Thiel, Richter, Liberati, Heeg, R{\"o}ckerath,
  Kourkoutis, M{\"u}hlbauer, B{\"o}ni, Muller, Barash, Schubert, Idzerda, Mannhart, and Schlom}}]{EuOSchlom}
\bibinfo{author}{\bibfnamefont{A.}~\bibnamefont{Schmehl}},
  \bibinfo{author}{\bibfnamefont{V.}~\bibnamefont{Vaithyanathan}},
  \bibinfo{author}{\bibfnamefont{A.}~\bibnamefont{Herrnberger}},
  \bibinfo{author}{\bibfnamefont{S.}~\bibnamefont{Thiel}},
  \bibinfo{author}{\bibfnamefont{C.}~\bibnamefont{Richter}},
  \bibinfo{author}{\bibfnamefont{M.}~\bibnamefont{Liberati}},
  \bibinfo{author}{\bibfnamefont{T.}~\bibnamefont{Heeg}},
  \bibinfo{author}{\bibfnamefont{M.}~\bibnamefont{R{\"o}ckerath}},
  \bibinfo{author}{\bibfnamefont{L.~F.} \bibnamefont{Kourkoutis}},
  \bibinfo{author}{\bibfnamefont{S.}~\bibnamefont{M{\"u}hlbauer}},
  \bibinfo{author}{\bibfnamefont{P.}~\bibnamefont{B{\"o}ni}},
  \bibinfo{author}{\bibfnamefont{D.~A.}~\bibnamefont{Muller}},
  \bibinfo{author}{\bibfnamefont{Y.}~\bibnamefont{Barash}},
  \bibinfo{author}{\bibfnamefont{J.}~\bibnamefont{Schubert}},
  \bibinfo{author}{\bibfnamefont{Y.}~\bibnamefont{Idzerda}},
  \bibinfo{author}{\bibfnamefont{J.}~\bibnamefont{Mannhart}}, \bibnamefont{and}
  \bibinfo{author}{\bibfnamefont{D.~G.}~\bibnamefont{Schlom}},
  \bibinfo{journal}{Nature Materials}
  \textbf{\bibinfo{volume}{6}}, \bibinfo{pages}{882} (\bibinfo{year}{2007}).

\bibitem[{\citenamefont{MacDonald et~al.}(2005)\citenamefont{MacDonald,
  Schiffer, and N.~Samarth}}]{MacDonaldDMS}
\bibinfo{author}{\bibfnamefont{A.~H.} \bibnamefont{MacDonald}},
  \bibinfo{author}{\bibfnamefont{P.}~\bibnamefont{Schiffer}}, \bibnamefont{and}
  \bibinfo{author}{\bibfnamefont{N.}~\bibnamefont{N.~Samarth}},
  \bibinfo{journal}{Nature Materials} \textbf{\bibinfo{volume}{4}},
  \bibinfo{pages}{195} (\bibinfo{year}{2005}).

\bibitem[{\citenamefont{Hetterich et~al.}(2006)\citenamefont{Hetterich,
  L{\"o}ffler, Fallert, H{\"o}pcke, Burger, Passow, Li, Daniel, Ramadout,
  Lupaca-Schomber, Hetterich, Litvinov, Gerthsen, Klingshirn, and Kalt}}]{HetterichPss}
\bibinfo{author}{\bibfnamefont{M.}~\bibnamefont{Hetterich}},
  \bibinfo{author}{\bibfnamefont{W.}~\bibnamefont{L{\"o}ffler}},
  \bibinfo{author}{\bibfnamefont{J.}~\bibnamefont{Fallert}},
  \bibinfo{author}{\bibfnamefont{N.}~\bibnamefont{H{\"o}pcke}},
  \bibinfo{author}{\bibfnamefont{H.}~\bibnamefont{Burger}},
  \bibinfo{author}{\bibfnamefont{T.}~\bibnamefont{Passow}},
  \bibinfo{author}{\bibfnamefont{S.}~\bibnamefont{Li}},
  \bibinfo{author}{\bibfnamefont{B.}~\bibnamefont{Daniel}},
  \bibinfo{author}{\bibfnamefont{B.}~\bibnamefont{Ramadout}},
  \bibinfo{author}{\bibfnamefont{J.}~\bibnamefont{Lupaca-Schomber}},
  \bibinfo{author}{\bibfnamefont{J.}~\bibnamefont{Hetterich}},
  \bibinfo{author}{\bibfnamefont{D.}~\bibnamefont{Litvinov}},
  \bibinfo{author}{\bibfnamefont{D.}~\bibnamefont{Gerthsen}},
  \bibinfo{author}{\bibfnamefont{C.}~\bibnamefont{Klingshirn}}, \bibnamefont{and}
  \bibinfo{author}{\bibfnamefont{H.}~\bibnamefont{Kalt}},
  \bibinfo{journal}{phys. stat. sol. (b)}
  \textbf{\bibinfo{volume}{243}}, \bibinfo{pages}{3812} (\bibinfo{year}{2006}).

\bibitem[{\citenamefont{Yuan et~al.}(2002)\citenamefont{Yuan, Kardynal,
  Stevenson, Shields, Lobo, Cooper, Beattie, Ritchie, and
  Pepper}}]{ElectricalSinglePhotonSourceScience}
\bibinfo{author}{\bibfnamefont{Z.}~\bibnamefont{Yuan}},
  \bibinfo{author}{\bibfnamefont{B.~E.} \bibnamefont{Kardynal}},
  \bibinfo{author}{\bibfnamefont{R.~M.} \bibnamefont{Stevenson}},
  \bibinfo{author}{\bibfnamefont{A.~J.} \bibnamefont{Shields}},
  \bibinfo{author}{\bibfnamefont{C.~J.} \bibnamefont{Lobo}},
  \bibinfo{author}{\bibfnamefont{K.}~\bibnamefont{Cooper}},
  \bibinfo{author}{\bibfnamefont{N.~S.} \bibnamefont{Beattie}},
  \bibinfo{author}{\bibfnamefont{D.~A.} \bibnamefont{Ritchie}},
  \bibnamefont{and} \bibinfo{author}{\bibfnamefont{M.}~\bibnamefont{Pepper}},
  \bibinfo{journal}{Science} \textbf{\bibinfo{volume}{295}},
  \bibinfo{pages}{102} (\bibinfo{year}{2002}).

\bibitem[{\citenamefont{Furdyna}(1988)}]{FurdynaDMS}
\bibinfo{author}{\bibfnamefont{J.~K.} \bibnamefont{Furdyna}},
  \bibinfo{journal}{J. Appl. Phys.} \textbf{\bibinfo{volume}{64}},
  \bibinfo{pages}{R29} (\bibinfo{year}{1988}).

\bibitem[{\citenamefont{Passow et~al.}(2007)\citenamefont{Passow, Li,
  Fein{\"a}ugle, Vallaitis, Leuthold, Litvinov, Gerthsen, and
  Hetterich}}]{PassowQDgrowth}
\bibinfo{author}{\bibfnamefont{T.}~\bibnamefont{Passow}},
  \bibinfo{author}{\bibfnamefont{S.}~\bibnamefont{Li}},
  \bibinfo{author}{\bibfnamefont{P.}~\bibnamefont{Fein{\"a}ugle}},
  \bibinfo{author}{\bibfnamefont{T.}~\bibnamefont{Vallaitis}},
  \bibinfo{author}{\bibfnamefont{J.}~\bibnamefont{Leuthold}},
  \bibinfo{author}{\bibfnamefont{D.}~\bibnamefont{Litvinov}},
  \bibinfo{author}{\bibfnamefont{D.}~\bibnamefont{Gerthsen}}, \bibnamefont{and}
  \bibinfo{author}{\bibfnamefont{M.}~\bibnamefont{Hetterich}},
  \bibinfo{journal}{J. Appl. Phys.} \textbf{\bibinfo{volume}{102}},
  \bibinfo{pages}{073511} (\bibinfo{year}{2007}).

\bibitem{footnote1}

Furthermore, it was explicitly shown that transitions between the
spin subbands in the QD due to spin relaxation are negligible. This
was inferred from experiments with a reference sample not containing
the DMS, see W. L{\"o}ffler, D. Tr{\"o}ndle, J. Fallert, H. Kalt, D.
Litvinov, D. Gerthsen, J. Lupaca-Schomber, T. Passow, B. Daniel, J.
Kvietkova, M. Gr{\"u}n, C. Klingshirn, and M. Hetterich, Appl. Phys.
Lett. \textbf{88}, 062105 (2006).


\bibitem[{\citenamefont{Asshoff et~al.}(2009)\citenamefont{Asshoff,
  L{\"o}ffler, Zimmer, F{\"u}ser, Fl{\"u}gge, Kalt, and
  Hetterich}}]{AsshoffSpinLEDPulsed}
\bibinfo{author}{\bibfnamefont{P.}~\bibnamefont{Asshoff}},
  \bibinfo{author}{\bibfnamefont{W.}~\bibnamefont{L{\"o}ffler}},
  \bibinfo{author}{\bibfnamefont{J.}~\bibnamefont{Zimmer}},
  \bibinfo{author}{\bibfnamefont{H.}~\bibnamefont{F{\"u}ser}},
  \bibinfo{author}{\bibfnamefont{H.}~\bibnamefont{Fl{\"u}gge}},
  \bibinfo{author}{\bibfnamefont{H.}~\bibnamefont{Kalt}}, \bibnamefont{and}
  \bibinfo{author}{\bibfnamefont{M.}~\bibnamefont{Hetterich}},
  \bibinfo{journal}{Appl. Phys. Lett.} \textbf{\bibinfo{volume}{95}},
  \bibinfo{pages}{202105} (\bibinfo{year}{2009}).

\bibitem[{\citenamefont{Becker}(2005)}]{BeckerHicklBook}
\bibinfo{author}{\bibfnamefont{W.}~\bibnamefont{Becker}},
  \emph{\bibinfo{title}{Advanced time-correlated single photon counting
  techniques}} (\bibinfo{publisher}{Springer}, \bibinfo{address}{Berlin}, \bibinfo{year}{2005}).

\bibitem[{\citenamefont{Patel et~al.}(2010)\citenamefont{Patel, Bennett,
  Cooper, Atkinson, Nicoll, Ritchie, and
  Shields}}]{StarkShiftSinglePhotonSource}
\bibinfo{author}{\bibfnamefont{R.}~\bibnamefont{Patel}},
  \bibinfo{author}{\bibfnamefont{A.}~\bibnamefont{Bennett}},
  \bibinfo{author}{\bibfnamefont{K.}~\bibnamefont{Cooper}},
  \bibinfo{author}{\bibfnamefont{P.}~\bibnamefont{Atkinson}},
  \bibinfo{author}{\bibfnamefont{C.}~\bibnamefont{Nicoll}},
  \bibinfo{author}{\bibfnamefont{D.}~\bibnamefont{Ritchie}}, \bibnamefont{and}
  \bibinfo{author}{\bibfnamefont{A.}~\bibnamefont{Shields}},
  \bibinfo{journal}{Nanotechnology} \textbf{\bibinfo{volume}{21}},
  \bibinfo{pages}{274011} (\bibinfo{year}{2010}).

\end{thebibliography}

\end{document}